\documentstyle[prl,aps,twocolumn]{revtex}
\def\eq{\begin{equation}}
\def\ee{\end{equation}}
\def\eqa{\begin{eqnarray}}
\def\eea{\end{eqnarray}}
\def\ra{\rightarrow}

\def\s{\sigma}
\def\p{\partial}
\def\psib{\bar{\psi}}
\def\bb{\bar{b}}
\def\rb{\bar{\rho}}     
                              
\def\w{\omega}

\def\ep{\epsilon}
\def\bt{\beta}
\def\ap{\alpha}

\begin{document}
\draft
\flushbottom
\twocolumn[
\hsize\textwidth\columnwidth\hsize\csname @twocolumnfalse\endcsname

\title{The Universal Composite Fermion Hall Conductance at $\nu=1/2$}
\author{D-H Lee$^{(a)}$, Y. Krotov$^{(a)}$, J. Gan$^{(a)}$ and S.A. Kivelson$^{(a,b)}$}
\address{$^{(a)}$
Department of Physics, University of California at Berkeley, 
Berkeley, CA 94720}
\address{$^{(b)}$Dept. of Physics
University of California at Los Angeles
Los Angeles, CA 90095} 
\date{\today}
\maketitle
\tightenlines
\widetext
\advance\leftskip by 57pt
\advance\rightskip by 57pt

\begin{abstract}
We show that at electronic filling factor $\nu=1/2$,
the lowest Landau level constraint implies that at any temperature,
and in the presence of {\it any} amount of particle-hole symmetric 
disorder, the {\it composite fermion} Hall conductivity is precisely $-e^2/2h$.
(The electronic Hall conductivity is $e^2/2h$.) 
This is inconsistent with the response of a composite Fermi liquid in 
{\it zero} effective magnetic field. We also examine perturbatively the nature of the 
putative composite Fermi liquid in the case in which the bare particles are 
anyons with ``nearly'' Fermi statistics.

\end{abstract}

\vskip 1cm
\pacs{73.50.Jt, 05.30.-d, 74.20.-z}

]

\narrowtext
\tightenlines

The observation of a seemingly metallic DC magnetotransport\cite{jiang} and
the subsequent discovery of an acoustic wave anomaly\cite{willet} near 
$\nu=1/2$, opened a new chapter in the studies of quantum Hall effects. 
(Here $\nu\equiv \phi_0\rb/B$, where $\rb$ is the mean electron density, 
$\phi_0 =hc/e$, and $B$ is the externally applied magnetic field.) 
A {\it very} intriguing idea, the composite fermion theory, has been put forward to 
explain these phenomena.\cite{hlr,kz}
In this theory, each electron is represented as a 
composite-fermion\cite{jain} carrying two quanta of fictitious magnetic flux 
which pierce the physical plane in the direction opposite to that of the real 
magnetic flux. 
Formally, this transformation maps the problem of electrons in a
strong magnetic field onto a system of ``composite
fermions'' moving in the same external field while interacting with a fluctuating ``statistical'' gauge field governed by 
a Chern-Simons action.\cite{zhk} 
%
%
At the mean-field level, the averaged statistical magnetic field, 
$\bb=2\phi_0\rb=|B|$, cancels the 
external one, and the composite-fermions see no net field. 
It has been argued that the transport properties  of the electrons near 
$\nu=1/2$ simply reflect the underlying Fermi liquid (or, possibly, the marginal 
Fermi liquid) behavior of the composite fermions in {\it zero} magnetic field. 
This intriguing picture acquired further support when
%
%
Fermi-surface-like features were observed in recent experiments.\cite{fsurf}

Despite these successes, there are concerns about the composite Fermi liquid 
theory. In part, these stem from the fact that when one tries to improve upon 
the mean-field theory(MFT) by including the fluctuations of the statistical magnetic 
field, one encounters divergences.\cite{hlr} Attempts to sum these 
divergences have led to suggestive, but so far inconclusive results.
\cite{stern,wilczek,marsden,ioffe,stamp}
%
%
The purpose of the present paper is to reexamine
the Fermi liquid picture when the cyclotron frequency, $\w_c=eB/mc$,  
is so large that the low energy states of the electrons lie primarily in the lowest Landau level. Our findings are 
as follows:
 
1. At $\nu=1/2$, and in the presence of particle-hole (p-h) symmetric 
external potentials,\cite{note} the projected Hamiltonian in the 
lowest Landau level has p-h symmetry.
%
%
In the limit 
$\w_c\ra\infty$, and {\it assuming} that this symmetry 
is not spontaneously broken, 
we have  shown that the electron Hall conductivity is temperature independent, 
with value $\s_{xy}=e^2/2h$.
So long as $\rho_{xx}\ne 0$, this result further
implies that the composite fermion Hall conductivity is
\eq
\s_{xy}^{f}=-e^2/2h.
\ee
%
%
We take Eq.(1) as implying the inadequacy of a {\it zero field} 
composite Fermi 
liquid in describing the system.
%

For finite $\w_c$ and in the absence of disorder, we
consider the more general problem in which the bare particles have fractional
statistics, $\theta$, defined so that they can be viewed as  composite fermions with 
$\theta$ fictitious flux quanta tied to each.\cite{hlr}  (See Eqs.(3) and (4))
Within this class of models, 
the problem of physical interest corresponds to $\theta=2$, while 
the problem is simple in the limit $\theta \ra 0$ with $\ap\equiv\hbar\w_c/E_c$ fixed. (Here $E_c=e^2/2\epsilon\sqrt{{\rb / \pi}}$, is the typical strength of the Coulomb interaction.)
In that limit, the composite fermions are the bare electrons, and MFT is exact.\cite{stern} 
Here, we consider only the case in which the magnetic field satisfies the
commensurability condition $B=\rb \theta /\phi_0$, so that
the net effective magnetic field seen by 
the composite fermions is zero.
Therefore aside from the
overall energy scale $E_c$, the only independent
parameters in the problem are $\theta$ and $\ap$. (Lowest Landau-level projection
means taking the limit $\ap\rightarrow \infty$.)
For $\ap >0$, and $\theta \ra 0$, we have computed  corrections to the MFT, and the results are as follows.

2. To the lowest order in $\theta$ the perturbative contributions (see Fig.1) to $\lim_{q\ra 0}\s_{xy}^f(q)$\cite{sxy} is {\it zero}. Thus the fact that the composite fermion system lacks time-reversal symmetry (Eq.(1)) is not manifested to O($\theta$). 
%

3. It is well known\cite{hlr} that to lowest order in $\theta$, the 
correction to the composite fermion self-energy due to {\it longitudinal} 
gauge fluctuations diverges logarithmically 
with the size of the system. It was pointed out\cite{hlr,stern} that this
arises
from virtual excitations across the energy gap $\hbar \omega_c$.  
We have shown that to the same order, and for {\it any} values of $\vec{k}$
and $\omega$, there are no divergent contributions to the
density-density correlation function.
We tentatively interpret this result as implying a confinement phenomenon:
only excitations which are ``statistical charge neutral'', such as particle-hole excitations, are physical,
while single composite 
fermion excitations are {\it not} part of the physical spectrum!
%

4. It is also well known\cite{hlr} that there are singular corrections to the
composite fermion self-energy in the limit $\omega \ra 0$ and 
$\mid \vec{k}\mid\ra k_F$ due to {\it transverse} gauge fluctuations, and these have been the focus
of much of the work in the field.\cite{stern,wilczek,marsden,ioffe,stamp}
However, if the single-composite fermion
is unphysical, the question arises whether this singularity appears in any physical correlation function. Strikingly, Kim {\it et al} showed that in an $1/N$ expansion, 
there is no singular corrections to the density-density and current-current correlation in the long wave-length limit.\cite{kim}
There do, however, remain singular corrections to the $\mid\vec{q}\mid=2k_F$ 
density-density 
correlation function due to the {\it transverse} gauge 
fluctuations.\cite{ioffe,stamp} To the lowest order in $\ap\theta$, we 
have found that
\eq
\Delta\Pi (\omega, 2k_F)/
\Delta\Pi_0 (\omega,2k_F)
=1+\theta\alpha C_1 ln\left[{E_F\over {\mid \omega\mid}}\right]. 
\label{eq:DeltaPi}
\ee
Here $\Pi(\w,q)\equiv\int d^2x dt e^{i(\w t-\vec{q}\cdot\vec{x})}
<\rho(x,t)\rho(0,0)>$ with 
$\Pi_0$ being the density-density correlation function of free electrons,
$\Delta\Pi(\w,q)=\Pi(\w,q)-\Pi(0,q)$, 
and $C_1 =-{1\over {\pi}}[{1\over 4}+ln({{\alpha\theta}\over 2})]$.
The above result shows 
that a small 
$\ap$, hence {\it strong} Landau level mixing, 
suppresses the 
amplitude of the singular corrections.

In sum, at finite $\w_c$, while the perturbative results (\ref{eq:DeltaPi})) cast doubts on the validity of a composite Fermi liquid description, they also suggest that observation of this breakdown may be difficult not only because i) the lack of time reversal symmetry of composite fermions does not show up in low order perturbation theory, but also due to:
ii) The density-density and current-current 
correlation functions do not show any anomalous behavior for 
$q\ne 2k_F$.\cite{ioffe,stamp,kim} iii) At $q=2k_F$ the anomaly in the density-density correlation function is a very weak one\cite{ioffe,stamp} (see Eq.(\ref{eq:DeltaPi})).
For $\theta=2$ (i.e. the case of physical interest), Eq.(1) implies that the 
composite fermions can not form a Fermi liquid at $\ap\ra\infty$. 
{\it If} the system is a composite Fermi liquid for small $\ap$, 
there must be a quantum phase transition at a critical $\ap_c$.
In the following we elucidate on some technical details, and expand
on several of the results:

{\bf The Model:}  In units where $\hbar=k_B=1$, the Euclidean 
Lagrangian for the composite fermions is given
by

\eqa
L[\psib,\psi,a]&=&\int d^2x\psib (\p_0-ieA_0+ia_0)\psi 
	\label{eq:L} \\
 -  &{1\over {2m}}& \int d^2x \psib({\vec{\nabla}}- 
i\frac e c
\vec{A}+i\vec{a})^2\psi +
L_{a}[a], \nonumber
\eea
where
\eqa
L_{a} &=& -{i\over 4\pi\theta}
\int d^2x
\ep^{\mu\nu\lambda}a_{\mu}\p_{\nu}a_{\lambda} 
\label{eq:Lga} \\
& + & {1\over 8\pi^2\theta^2}\int d^2xd^2x'
[ b(x,t)-\bb ] V(x-x')[
b(x',t)-\bb ]. \nonumber 
\eea
In the above $\psib$ and $\psi$ are the Grassmann fields associated with the 
composite fermions; $A_{\mu}$ and $a_{\mu}$ are the external and statistical 
gauge fields respectively; $m$ is the electron bare effective mass; 
$b=\vec{\nabla}\times \vec{a}$; $V(x-x')$ is the bare 
interaction between electrons; $\bb\equiv 2\pi\theta\rb$ is the averaged statistical magnetic field. 
Moreover, we have made use of the Chern-Simons constraint that
$b(x,t)=2\pi\theta \rho(x,t)$.
By rescaling space, time, and the fermion fields, so that 
$x\ra k_F x, t\ra tk_F^2/m$, and $\psi,\psib\ra k_F^{-1}\psi, 
k_F^{-1}\psib$, (here 
$k_F\equiv\sqrt{\rb/\pi}$)
one can easily prove that in Eq.(3) and (4) the only dimensionless 
parameters are $\theta$ and $\ap$.

An important ingredient of the above Chern-Simons formulation is the 
relation between the bare particle, and composite fermion correlation 
functions.  It is the nature of the mapping that the density
of composite fermions equals that of the bare particles,  but the 
relation between current operators is more complicated.  To compute 
the 
bare particle current-current correlation function, we need to string 
together the composite fermion ``irreducible bubbles''\cite{klz} using 
the bare gauge propagator $<a_{\mu} a_{\nu}>$ computed from Eq.(4). 
As shown in Refs.\cite{hlr,klz} this results in the following relations 
between the resistivity tensor of bare particle $\rho_{\ap\bt}$ and 
that of the composite fermion $\rho_{\ap\bt}^f$: 
\eqa
&\rho_{xx}&=\rho_{xx}^{f} \nonumber \\
&\rho_{yx}&=\theta {h\over {e^2}}+\rho_{yx}^{f}.
\eea
Physically, this expresses the fact that associated with the
composite fermion current, there is a statistical flux current,
which produces a corresponding EMF proportional to $\theta$ times
the electrical current.

In Ref.\cite{stern} Stern and Halperin 
defined the ``(marginal) composite Fermi liquid'' by requiring the 
irreducible bubbles to be that of a (marginal) Fermi liquid in 
{\it zero} magnetic field. Among other things, 
this assertion implies that 
$\s_{xy}^f=0$. For subsequent discussions, we take this as a working 
definition 
of a (marginal) composite Fermi liquid.  
This is a reasonable definition since, after all, the idea of the 
composite 
fermion approach is to describe the behavior of electrons at $\nu=1/2$ 
in terms of those of composite fermions in {\it zero} field.

{\bf  Lowest Landau Level Projection:}
For the case of physical interest, $\theta=2$, and in the
limit $\ap\ra\infty$, the low energy eigenstates are solely made up of 
states in the lowest Landau level.
The projection of the electron Hamiltonian into the lowest Landau level 
gives $
H_L^0=\mu\int d^2x\rho_L(x)+{1\over 2}
\int d^2xd^2x'
V(x-x'){:\rho_L(x)\rho_L(x'):}$.
Here $\rho_L(x)=\psi_L^+(x)\psi_L(x)$, with $\psi_L\equiv\sum_k\psi_k(x)c_k$,
where $\psi_k$ is the lowest Landau level basis and $c_k$ is the associated annihilation operator.
The p-h transformation is implemented via
$\psi_L(x)\ra\psi_L^+(x)=\sum_{k}\psi^*_k(x)c^+_k$.

At $\nu=1/2$, the value of $\mu$ is such that $H_L$ is
invariant under the p-h transformation. In the presence of disorder potential $H_L=H_L^0+\int d^2x U(x)\psi_L^+(x)\psi_L(x)$. Thus, 
$H_L[U(x)]\ra H_L[-U(x)]$ under p-h transformation. To prove that $\s_{xy}=e^2/2h$ when $\w_c\ra\infty$, we start with the Kubo formula
$\s_{xy}(\w)={1\over \w}\int dt e^{i\w t}\theta(t)<[j_x(t),j_y(0)]>$, 
where $<...>$ stands for the quantum, thermal, and impurity averages,
$j_{\ap}\equiv {1\over A}\int d^2x j_{\ap}(x)$  
($A$ is the total area). Then we let $\w_c\ra\infty$ and keep all terms to order $(1/\w_c)^0$.\cite{sd} Finally we perform the 
p-h transformation. The details will be reported elsewhere but the 
result is self-evident. {\it If} there is no spontaneous particle-hole 
symmetry breaking,
\eq
\s_{xy}\equiv \lim_{\w\ra 0}
\s_{xy}(\w)=-\s_{xy}+e^2/h.
\label{eq:12}
\ee
In the above,
$e^2/h$ arises as the
Hall conductivity of the hole vacuum, i.e., 
the filled lowest Landau level. Eq.(\ref{eq:12}) 
implies that $\s_{xy}=e^2/2h$!

So long as $\rho_{xx}\ne 0$, it can be seen directly from Eq.(5) 
that $\s_{xy}=e^2/2h$ implies $\s_{xy}^f=-e^2/2h$. In the literature it 
is noted that 
potential disorder for the electrons induces both potential and 
magnetic flux disorder for the composite fermions.\cite{hlr,kz} 
According to this picture, when the potential disorder is p-h symmetric, 
the associated random magnetic field is also symmetrically
distributed about zero. Under that 
condition, the {\it impurity-averaged} Hall conductivity $\s_{xy}^f$ 
should vanish, in contradiction with the requirements 
imposed by the projection onto the lowest Landau level.

{\bf Perturbative Results:}
When the flux and particle density are related according to 
$B=\theta\phi_0\rb$,
%
%
the {\it average} statistical magnetic 
field seen by the composite fermions
exactly cancels the external one.
%
%
The question remains, what 
are the fluctuation corrections to this mean-field picture?  
As mentioned above,
in the limit $\theta\rightarrow 0$ with non-zero $\ap$, 
fluctuations of the statistical gauge field trivially vanish,
%
%
%
and this limit provides a reference point where MFT is 
exact.\cite{stern}
In carrying out these fluctuation calculations, we choose
to work in Coulomb gauge, in which the gauge-field propagator, $D_{ij}$,
is a $2\times 2$ matrix, with $j=0, 1$ representing the time and space
component, respectively.

First, we comment on our result concerning $\lim_{q\ra 0}\s_{xy}^f(q)$. 
We calculated $\s_{xy}^f$ perturbatively by evaluating the 
Feynman diagrams shown in Fig.1. In 
that figure the wavy line represents the mixed gauge propagator $D_{01}$ 
and $D_{10}$. The open, solid triangles, and the square represent the density, current and the diamagnetic vertices respectively. 
To the lowest order in $\theta$ and $\ap$ we used the bare gauge propagator. In this case, since $D_{01}$ does not depend on frequency, 
the integration can be easily done and the contributions to $\s_{xy}^f$ from Figs.1(a), 1(b), and 1(c),1(d) are $\pm\theta/8(e^2/h)$ respectively, thus the net result is zero.
In this calculation we found that the characteristic momentum carried by 
the gauge line is of order $k_{F}$.  

Next, we summarize our results for the density-density correlation 
function. 
The bare gauge propagator in the Coulomb gauge has the property that $D_{11}=0$ 
and $D_{00}(q_0,\vec{q})=V(\vec{q})$. Thus the effects of $D_{00}$ are 
identical to those of a static Coulomb interaction. 
As is customary in this case, a RPA resumation is
performed to 
screen $D_{00}$ and $D_{11}$. If one uses the renormalized $D_{00}$ 
and $D_{11}$ to compute the 1-loop corrections to the composite fermion self-energy, $\Sigma(q_0,\vec{q})$, 
the contribution from longitudinal fluctuations, {\it i.e.} those
which involve $D_{00}$, diverges logaritmically with the
size of the system for fixed $q_0$ and $\vec{q}$, and
the contribution from transverse fluctuations, {\it i.e.}
those involving  $D_{11}$, are regular in the system size, but
contribute a logarithmicly diverging correction to the 
effective mass.\cite{hlr} However, at the same level of 
approximation in computing the density-density correlation function 
(see Fig.2), we find that the singular self-energy correction caused by 
$D_{00}$ is canceled by the corresponding vertex correction for 
{\it all} $\vec{q}$ and $\w$. To prove this we have
used the fact that the 
{\it divergent part} of the vertex $
\Gamma$ and the self-energy are related by the following identity
\begin{equation}
\Gamma(p,p+q) = \frac{\Sigma(p) - \Sigma(p+q)}{iq_{0} - \epsilon(p+q) + \epsilon(p)},
\end{equation}
where $\vec{q}$ and $q_{0}$ are the incoming gauge field momentum and 
frequency. After summing a)-c) in Fig.2 the {\it divergent part} of the 
density-density correlation function is given by
\begin{equation}
\int \frac{d^{3}p}{(2 \pi)^{3}} \frac{[G(p+q)]^{2} 
\Sigma(p+q) - [G(p)]^{2} \Sigma(p)}{iq_{0} - \epsilon(p+q) + 
\epsilon(p)},
\end{equation} 
which vanishes after integration over $p_{0}$ because the poles of 
$G$ (the composite fermion propagator)
and  $\Sigma$ lie on the same side from the real axis. 
The same thing can not be said for the corrections caused by $D_{11}$. 
In that case the singular contribution from the self-energy and vertex 
corrections do not cancel for $\mid\vec{q}\mid=2k_F$. 
(They do cancel at other $\mid\vec{q}\mid$.\cite{ioffe,stamp,kim}) The graphs used in that 
calculation are summarized in Fig.2. The result for the $2k_F$ 
density-density correlation function is given by Eq.(2).

The fact that the divergent self-energy correction from $D_{00}$ is 
canceled by the vertex correction for {\it all} external momenta sheds light on 
the following fundamental issue.  
The divergent self-energy correction caused by $D_{00}$ stems from the 
pole of $D_{00}$ at $\w=\w_c$. It is eliminated if 
one suppresses all {\it intermediate states} that are not 
in the lowest Landau level. This has been taken as an indication that 
the Landau level projection is, somehow, a {\it necessary} step in constructing a 
meaningful theory for composite fermions.\cite{hlr,stern} Our 
result suggests an alternative view. We believe that single composite 
fermion excitations are {\it not} part of the physical spectrum. 
Instead, the physical excitations are {\it statistical 
charge-neutral} p-h 
excitations.\cite{kim} Thus {\it to describe physical excitations, the 
Landau level projection is not necessary.} 
Of course, we have proven the consistency of this viewpoint only 
to lowest order in perturbation theory, so at this point we can only
conjecture that it remains valid more generally.

Our result for $\s_{xy}^f$ is strictly perturbative 
in $\theta$. The same is not true of the result for
$\Pi(\w,\vec{q})$, where  
a RPA resumation has been performed. If we take Eq.(2) at 
face value, there will be a crossover temperature/frequency below which 
the singular corrections to $\Pi$ at $\mid\vec{q}\mid=2k_F$ becomes 
significant.
%
 
{\bf Possible relevance to experiment:}
Our most important conclusion is that when a Hall sample exhibits p-h symmetry at long wave-length, i.e. $\s_{xy}=e^2/2h$, it is {\it not} describable
as a composite Fermi liquid.  In a recent study\cite{hwj}
of gated GaAs heterojunctions with mobility
$\mu \le 2\times 10^6 cm^2/Vs$, a line in the density-magnetic field plane has been identified at which $\nu \approx 1/2$, and $\sigma_{xy}=e^2/2h$ independent of temperature from 50mK to 1.5K. (Meanwhile $\rho_{xx}$ varies
with temperature and density, taking
values between $0.02$ to $1 h/e^2$.) We interpret this result as indicating
that p-h symmetry is respected in real systems near $\nu=1/2$. This line merges with the phase boundary between the $\nu=1$ and the insulating phase 
(on which $\sigma_{xx}=\sigma_{xy}=e^2/2h$ at low temperatures).
In the same experiments, a line is also observed on which
$\rho_{xy}\approx 2h/e^2$, and is approximately temperature independent. (These two lines converge as $\rho_{xx}\ra 0$.)  If the composite Fermi liquid exists, it must
be along this latter line. Finally, 
our perturbative analyses suggest that there can exist
a crossover temperature which, according to Eq. (2), is exponentially small in the limit of large Landau level mixing, which separates a high temperature regime in which the mean-field Fermi liquid theory is valid, from a low temperature regime in which more subtle fluctuation physics pertains.  (A similar observation was made previously in Ref. \cite{ioffe}.)  

The present considerations do not directly address the  nature of the true composite fermion ground state at $\nu=1/2$ in the limit of $\ap\ra\infty$.  
It is possible that this state possesses a Fermi surface, and 
has $\sigma_{xy}^f=-e^2/2h$.

{\bf Acknowledgments:}  We acknowledge illuminating
discussions with E.~Fradkin, L.~Ioffe, A.~C.~Neto, P. Stamp, 
and X-G.~Wen.  We thank Dr. D. Khveshchenko for pointing out an error in our earlier result for $lim_{q\ra 0}\s_{xy}^f(q)$. SK was
supported in part by the NSF under grant number DMR93-12606 at UCLA, and
by a Miller Fellowship at UCB.

\centerline{\bf Figure Captions}

\ \

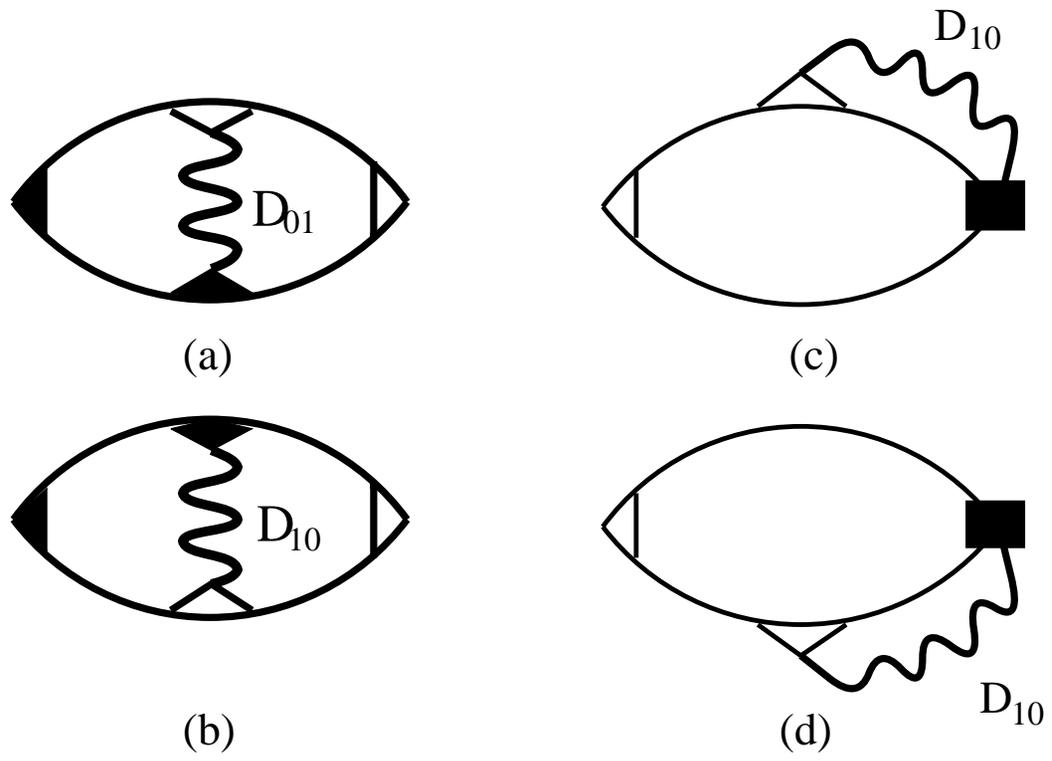
\noindent{\bf Fig.~1}. Feynman diagrams for $lim_{q\ra 0}\s_{xy}^f (q)$. Note that the diagrams a),b) cancels diagram c), d). Moreover, the diagrams corresponding to self-energy insertions vanish due to symmetry.
\ \

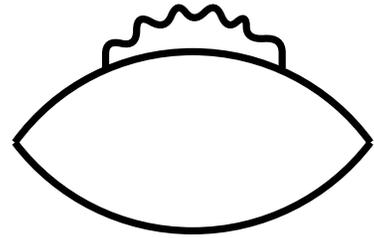
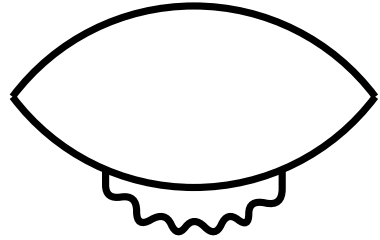
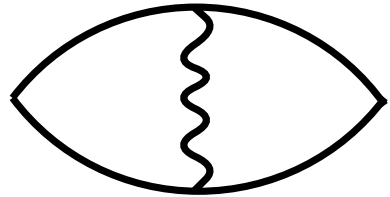
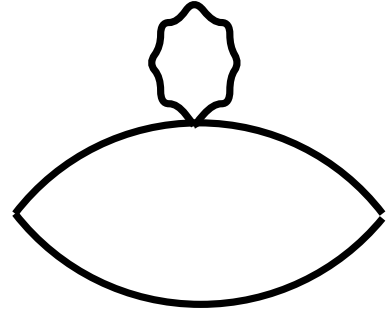
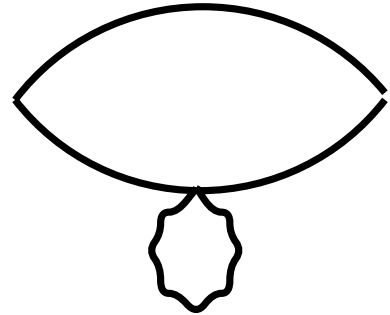
\noindent{\bf Fig.~2}. Feynman diagrams for $\Pi(q_0,\vec{q})$. For logitudinal gauge fluctuations, diagrams d) and e) are absent.
\vspace*{\fill}

\begin{references}
\bibitem{jiang} H.W.Jiang et al., Phys.Rev.Lett. {\bf65}, 633 (1990). 
\bibitem{willet} R.L.Willet et al.,Phys.Rev.Lett. {\bf71}, 3846 (1993). 
\bibitem{hlr}  B. I. Halperin, P. A. Lee, and N. Read, Phys. Rev. B {\bf47}, 7312 (1993).
\bibitem{kz} V.Kalmeyer and S.C.Zhang, Phys.Rev.B {\bf46}, 9889 (1992). 
\bibitem{jain} J.K.Jain, Phys.Rev.Lett. {\bf63}, 199 (1989).
\bibitem{zhk} S.C. Zhang, H. Hanson and S. Kivelson, Phys. Rev. Lett. 
{\bf62}, 82 (1989); {\bf62}, 980 (E) (1989).  A.~Lopez and E.~Fradkin,
Phys. Rev. B {\bf 44}, 5246 (1991). 
\bibitem{fsurf} W.Kang et al., Phys.Rev.Lett. {\bf71}, 3846 (1993);
V.J. Goldman, B.Su and J.K. Jain, Phys. Rev. Lett. {\bf72}, 2065 (1994); 
R.L.Willet, K.W.West and L.N.Pfeiffer, Phys.Rev.Lett. {\bf75}, 2988 (1995).
\bibitem{stern} A. Stern and B.I. Halperin, Phys. Rev. B {\bf52}, 5890 (1995).
\bibitem{wilczek} C. Nayak and F. Wilczek, Nucl.Phys.B {\bf417}, 359 (1994).
\bibitem{marsden} H-J Kwon, A. Houghton and J.B. Marston, Phys. Rev. B {\bf52}, 8002 (1995).
\bibitem{ioffe} B.L. Altshuler, L.B. Ioffe and A.J. Millis, Phys. Rev. B {\bf50}, 14048 (1994).
\bibitem{stamp} D.V. Khveshchenko and P.C.E.Stamp, Phys.Rev.Lett. {\bf71}, 2118 (1993); Phys.Rev.B {\bf49}, 5227 (1994).
\bibitem{note} In the presence of disorder potential $U(x)$, 
we speak of p-h symmetry if the disorder ensemble has the property that $P[U(x)]=P[-U(x)]$, where $P[U]$ is the probability that a particular $U(x)$ is realized. Here,
without loss of generality, we have set $\int d^2x U(x)=0$.
\bibitem{sxy}  This is not the cannonical order of limits, but for the purpose of illustrating the lack of time reversal symmetry it is sufficient.  
\bibitem{kim} Y.B. Kim, A. Furusaki, X-G Wen and P.A. Lee, Phys. Rev. B {\bf50}, 17917 (1994).
\bibitem{klz} S.Kivelson, D.H.Lee and S.C.Zhang, Phys.Rev.B {\bf46}, 2223 (1992).
\bibitem{sd} R.Rajaraman and S.L.Sondhi, Int. J. Mod. Phys. B {\bf8}, 1065 (1994).

\bibitem{hwj}L.~W.Wong and H.W. Jiang, unpublished.
\end{references}
\end{document}